\documentclass[aps,prl,reprint,superscriptaddress,showpacs]{revtex4-1}

\usepackage{graphicx}
\usepackage{color}
\usepackage{amsfonts}

\begin{document}
\title{Spectroscopic evidence for polaronic behaviour of the strong spin-orbit insulator Sr$_3$Ir$_2$O$_7$} 

\author{P.~D.~C.~King}
\affiliation{SUPA, School of Physics and Astronomy, University of St. Andrews, St. Andrews, Fife KY16 9SS, United Kingdom}
\affiliation{Kavli Institute at Cornell for Nanoscale Science, Ithaca, New York 14853, USA}
\affiliation{Laboratory of Atomic and Solid State Physics, Department of Physics,
Cornell University, Ithaca, New York 14853, USA}

\author{T.~Takayama}
\affiliation{Department of Physics, University of Tokyo, Hongo, Tokyo 113-0033}

\author{A.~Tamai}
\affiliation{SUPA, School of Physics and Astronomy, University of St. Andrews, St. Andrews, Fife KY16 9SS, United Kingdom}
\affiliation{D{\'e}partement de Physique de la Mati{\`e}re Condens{\'e}e, Universit{\'e} de Gen{\`e}ve, 24 Quai Ernest-Ansermet, 1211 Gen{\`e}ve 4, Switzerland}
\author{E.~Rozbicki}
\affiliation{SUPA, School of Physics and Astronomy, University of St. Andrews, St. Andrews, Fife KY16 9SS, United Kingdom}
\author{S.~McKeown Walker}
\affiliation{SUPA, School of Physics and Astronomy, University of St. Andrews, St. Andrews, Fife KY16 9SS, United Kingdom}
\affiliation{D{\'e}partement de Physique de la Mati{\`e}re Condens{\'e}e, Universit{\'e} de Gen{\`e}ve, 24 Quai Ernest-Ansermet, 1211 Gen{\`e}ve 4, Switzerland}

\author{M.~Shi}
\affiliation{Swiss Light Source, Paul Scherrer Institut, CH-5232 Villigen PSI, Switzerland}
\author{L.~Patthey}
\affiliation{Swiss Light Source, Paul Scherrer Institut, CH-5232 Villigen PSI, Switzerland}
\affiliation{SwissFEL, Paul Scherrer Institut, CH-5232 Villigen PSI, Switzerland}

\author{R.~G.~Moore}
\author{D.~Lu}
\affiliation{Stanford Institute for Materials and Energy Sciences, SLAC National Accelerator Laboratory, 2575 Sand Hill Road, Menlo Park, California 94025, USA}
\affiliation{Geballe Laboratory for Advanced Materials, Departments of Physics and Applied Physics, Stanford University, California 94305, USA}

\author{K.~M.~Shen}
\affiliation{Kavli Institute at Cornell for Nanoscale Science, Ithaca, New York 14853, USA}
\affiliation{Laboratory of Atomic and Solid State Physics, Department of Physics,
Cornell University, Ithaca, New York 14853, USA}

\author{H.~Takagi}
\affiliation{Department of Physics, University of Tokyo, Hongo, Tokyo 113-0033}
\affiliation{Magnetic Materials Laboratory, RIKEN Advanced Science Institute, Wako, Saitama 351-0198, Japan}

\author{F.~Baumberger}
\affiliation{SUPA, School of Physics and Astronomy, University of St. Andrews, St. Andrews, Fife KY16 9SS, United Kingdom}
\affiliation{D{\'e}partement de Physique de la Mati{\`e}re Condens{\'e}e, Universit{\'e} de Gen{\`e}ve, 24 Quai Ernest-Ansermet, 1211 Gen{\`e}ve 4, Switzerland}
\affiliation{Swiss Light Source, Paul Scherrer Institut, CH-5232 Villigen PSI, Switzerland}

\date{\today}

\begin{abstract}\noindent We investigate the bilayer Ruddlesden-Popper iridate Sr$_3$Ir$_2$O$_7$ by temperature-dependent angle-resolved photoemission. We find a narrow-gap correlated insulator, with spectral features indicative of a polaronic ground state, strikingly similar to that observed previously for the parent compounds of the cuprate superconductors. We additionally observe similar behaviour for the single-layer cousin Sr$_2$IrO$_4$, indicating that strong electron-boson coupling dominates the low-energy excitations of this exotic family of materials, and providing a microscopic link between the insulating ground states of the seemingly-disparate $3d$ cuprates and $5d$ iridates. \end{abstract}


\maketitle

\begin{figure*}
\begin{center}
\includegraphics[width=\textwidth]{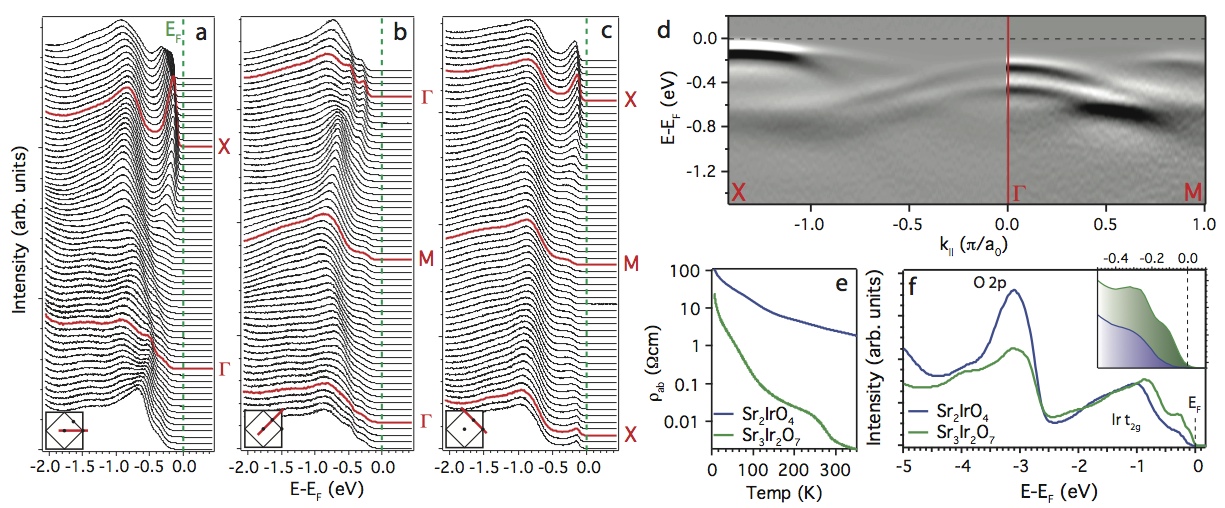}
\caption{ \label{f:overview} Low-energy electronic structure of Sr$_3$Ir$_2$O$_7$, measured with a photon energy of 40~eV at 50~K along the (a) $\Gamma$-$X$, (b) $\Gamma$-$M$, and (c) $M$-$X$ high-symmetry directions of the Brillouin zone (see insets). (d) Second-derivative image plot of the electronic structure along $X$-$\Gamma$-$M$. (e) Temperature-dependent resistivity and (f) valence band photoemission of the bilayer (green) and single-layer (blue) compounds. A magnified view of the near-$E_F$ valence band emission is shown inset in (f).}
\end{center}
\end{figure*}

The strong spin-orbit interaction in the $5d$ shell is predicted to stabilize a variety of exotic ground states in iridium-based transition-metal oxides, including Mott insulators~\cite{Kim:Phys.Rev.Lett.:101(2008)076402,Kim:Science:323(2009)1329--1332,Moon:Phys.Rev.Lett.:101(2008)226402}, Weyl semimetals~\cite{Wan:Phys.Rev.B:83(2011)205101}, correlated topological insulators~\cite{Shitade:Phys.Rev.Lett.:102(2009)256403,Pesin:NaturePhys.:6(2010)376--381,Xiao:NatureCommun.:2(2011)596--,Yang:Phys.Rev.B:82(2010)085111} and spin-triplet superconductors~\cite{You:arXiv:1109.4155:(2011)}. Moreover, iridates were recently proposed as an analogue of the cuprates, and as such, a potential platform to engineer high-temperature superconductivity~\cite{Wang:Phys.Rev.Lett.:106(2011)136402}. This initially appears surprising given the weak influence of electron correlations expected for spatially-extended $5d$ orbitals. Nonetheless, Sr$_2$IrO$_4$ and Sr$_3$Ir$_2$O$_7$, which both host partially-filled $5d$ shells, are found to be insulating~\cite{Crawford:Phys.Rev.B:49(1994)9198--9201,Cao:Phys.Rev.B:66(2002)214412}. For Sr$_2$IrO$_4$, this was recently attributed~\cite{Kim:Phys.Rev.Lett.:101(2008)076402,Kim:Science:323(2009)1329--1332} to a reconstruction of the underlying electronic structure by a co-operative interplay of structural distortions and, crucially, the strong spin-orbit coupling, leaving a half-filled $J_{eff}=1/2$ band that is sufficiently narrow that even moderate correlation strengths can drive a Mott transition. However, the range of validity of this strong spin-orbit $J_{eff}=1/2$ Mott picture remains an open question~\cite{Martins:Phys.Rev.Lett.:107(2011)266404,Liu:Phys.Rev.Lett.:109(2012)157401,Gretarsson:arXiv:1209.5424:(2012),Arita:Phys.Rev.Lett.:108(2012)086403,Hsieh:Phys.Rev.B:86(2012)035128}, as does the microscopic similarity of the insulating ground state to that of the parent compounds in the cuprates.

Indeed, kinetic, Coulomb, crystal-field, and spin-orbit energy scales are all of similar magnitude in the iridates, potentially leading to the close proximity of several competing ground states. For example, optical conductivity measurements revealed a metal-insulator transition (MIT) upon increasing dimensionality through the layered Ruddlesden-Popper series Sr$_{n+1}$Ir$_n$O$_{3n+1}$~\cite{Moon:Phys.Rev.Lett.:101(2008)226402}, with the conducting three-dimensional end member predicted to be an exotic semi-metal~\cite{Carter:Phys.Rev.B:85(2012)115105}. The intermediate $n=2$ bilayer compound is thought to lie close to the borderline between the (semi-)metallic and insulating phases~\cite{Moon:Phys.Rev.Lett.:101(2008)226402}. As such, it promises new insights on the nature of the MIT and the unconventional insulating ground states of iridates. 

Here, we study its low-energy electronic structure by angle-resolved photoemission (ARPES). We find a small charge gap to a weakly-dispersive band, reminiscent of a $J_{eff}=1/2$ lower Hubbard band, although with other nearby dispersive bands that complicate this picture. Moreover, we observe a pronounced temperature-dependent broadening of the photoemission linewidths, and spectral lineshapes notably similar to that observed in cuprates~\cite{Shen:Phys.Rev.Lett.:93(2004)267002} and manganites~\cite{Dessau:Phys.Rev.Lett.:81(1998)192--195}, and indicative of a polaronic ground state in layered $5d$ iridates.

ARPES measurements were performed at the SIS beamline of the Swiss Light Source, beamline V-4 of Stanford Synchrotron Radiation Lightsource, and using a laboratory He-lamp system. Measurements were made using photon energies between $16$ and $120$ eV and both Scienta R4000 and SPECS Phoibos 225 hemispherical analyzers. The sample temperature was varied between 25 and 350~K. Single-crystal samples of Sr$_3$Ir$_2$O$_7$ and Sr$_2$IrO$_4$ were flux-grown. These were cleaved {\it in-situ} at pressures better than $5\times10^{-11}$~mbar. 

\begin{figure}
\begin{center}
\includegraphics[width=\columnwidth]{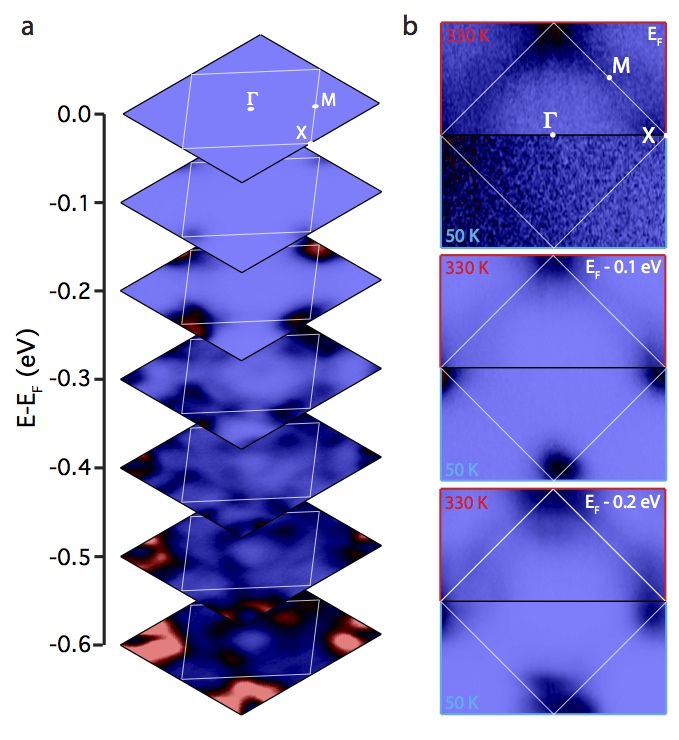}
\caption{ \label{f:FS} (a) Constant energy contours measured at 50~K and normalized to the same total intensity for each energy. (b) Comparison of low- (50~K) and high-temperature (330~K) maps at the Fermi level and 0.1 and 0.2~eV binding energy, normalized to the maximum intensity at that binding energy to maximize contrast.}
\end{center}
\end{figure}
The electronic structure of Sr$_3$Ir$_2$O$_7$ measured at 50~K along high-symmetry directions, as well as its momentum dependence across the full Brillouin zone, is summarized in Fig.~\ref{f:overview} and Fig.~\ref{f:FS}(a), respectively~\cite{Note1}. We find that no bands intersect the Fermi level, with spectral weight tending to zero by the chemical potential throughout the Brillouin zone. 
This identifies the low-temperature phase of Sr$_3$Ir$_2$O$_7$ as a fully-gapped insulator, consistent with the nearly exponential temperature dependence of its resistivity (Fig.~\ref{f:overview}(e)). However, comparison of angle-integrated spectra (Fig.~\ref{f:overview}(f)) to its single-layer counterpart reveals significantly-enhanced spectral weight close to the Fermi level in the bilayer compound, as well as a much smaller charge gap. 

Indeed, we observe dispersive features with well defined peaks centred as little as $\sim\!100$~meV below the Fermi level. The lowest-energy features are located at the X-point [$(\pi,0)$ point of the tetragonal Brillouin zone], as in the single-layer compound~\cite{Kim:Phys.Rev.Lett.:101(2008)076402}. This is in qualitative disagreement with LDA+U calculations, even when spin-orbit interactions are included~\cite{Moon:Phys.Rev.Lett.:101(2008)226402}, which predict that the top of the valence band occurs at the zone centre. Second-derivative plots (Fig.~\ref{f:overview}(d)) reveal that two bands disperse away from the band maximum at the X-point. We estimate that the upper band, which evolves into a weak shoulder near the M-point [$(\pi/2,\pi/2)$, Fig.~\ref{f:overview}(b,c)], disperses by only $\sim\!150-200$~meV across the Brillouin zone.

This narrow bandwidth appears broadly consistent with a very narrow $J_{eff}=1/2$ lower Hubbard band, as proposed for other iridates~\cite{Kim:Phys.Rev.Lett.:101(2008)076402,Comin:arXiv:1204.4471:(2012)}. In angle-integrated spectra of the single-layer compound (Fig.~\ref{f:overview}(f)), two dominant peaks at $\sim\!0.25$~eV and $\sim\!1$~eV are observed, which might be assigned as the origin of the $\alpha$ and $\beta$ transitions in optical conductivity~\cite{Moon:Phys.Rev.Lett.:101(2008)226402}. Such features, observed in several iridate oxides, have been attributed to optical transitions to the upper Hubbard band from a single $J_{eff}=1/2$ lower Hubbard band and a separate $J_{eff}=3/2$ band, respectively~\cite{Kim:Phys.Rev.Lett.:101(2008)076402,Moon:Phys.Rev.Lett.:101(2008)226402,Kuriyama:Appl.Phys.Lett.:96(2010)182103,Comin:arXiv:1204.4471:(2012)}. In contrast, the situation already appears more complex in angle-integrated spectra from Sr$_3$Ir$_2$O$_7$, with additional peaks present at low binding energies compared to Sr$_2$IrO$_4$. This could reflect the presence of a non-negligible octahedral crystal field splitting~\cite{Liu:Phys.Rev.Lett.:109(2012)157401}. Moreover, departures from an idealized $J_{eff}=1/2,3/2$ picture can be readily identified in angle-resolved spectra. For example, at the $X$ point, two well separated peaks are observed in energy distribution curves (EDCs), which could naturally be assigned as distinct $J_{eff}=1/2$ and $3/2$ manifolds. However, significant dispersion is observed at other momenta, which presumably leads to small avoided crossings that are difficult to resolve experimentally, as well as substantial mixing of the $J_{eff}=1/2$ and $3/2$ states. 

It is therefore an oversimplification to describe Sr$_3$Ir$_2$O$_7$ as a prototypical $J_{eff}=1/2$ Mott insulator. This raises the question whether the insulating state results from some form of multi-band Mott-like transition, or whether additional order parameters such as magnetism may play a crucial role~\cite{Arita:Phys.Rev.Lett.:108(2012)086403,Hsieh:Phys.Rev.B:86(2012)035128,Fujiyama:arXiv:1207.7151:(2012)}, for example within a Slater-type picture. Consistent with previous measurements~\cite{Fujiyama:arXiv:1207.7151:(2012)}, we observe a pronounced drop in the resistivity of the bilayer compound at $\sim\!280$~K (Fig.~\ref{f:overview}(e)), concurrent with magnetic ordering~\cite{Cao:Phys.Rev.B:66(2002)214412}. We find no sharp changes in the electronic structure measured by ARPES through this transition. Even at temperatures as high as 330~K (Fig.~\ref{f:FS}(b)), there is still only minimal spectral weight at the Fermi level. At energies below the Fermi level, the spectral weight across the Brillouin zone is dominated by the top of the hole-like bands at the $X$ point at both high (Fig.~\ref{f:FS}(b), top panels) and low (Fig.~\ref{f:FS}(b), bottom panels) temperatures. At the Fermi level itself, there appears to be a relatively-higher contribution from states around the $M$ point in the higher-temperature measurements. This could suggest a closing of the insulating charge gap from below $E_F$ at the $X$ point and above $E_F$ at the $M$ point, indicating a transition from an insulator to a semi-metal with increasing temperature. However, this manifests itself as a very gradual increase in spectral weight at the Fermi level, and does not appear to occur as a sudden phase transition at $T_N$.

\begin{figure}[!h]
\begin{center}
\includegraphics[width=\columnwidth]{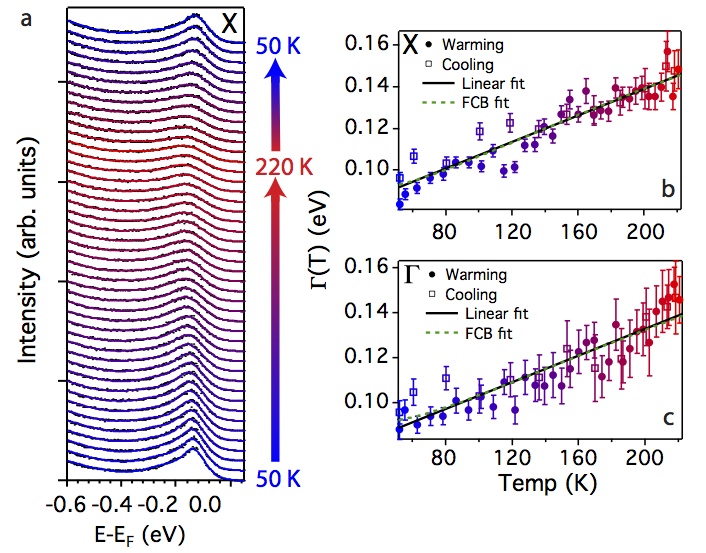}
\caption{ \label{f:T-dep} (a) Temperature-dependence of EDCs (black dots) and fits (to two gaussian peaks and a polynomial background, solid lines) at the X-point. The spectra were all measured on a single sample, from bottom to top with increasing time. The width, $\Gamma(T)$, of the lowest-energy peak is shown in (b), and for equivalent measurements at the $\Gamma$-point in (c). Black lines show linear fits to the full data set, green-dashed lines show a model fit for Franck-Condon broadening, discussed in the main text.}
\end{center}
\end{figure}
\begin{figure}[!h]
\begin{center}
\includegraphics[width=\columnwidth]{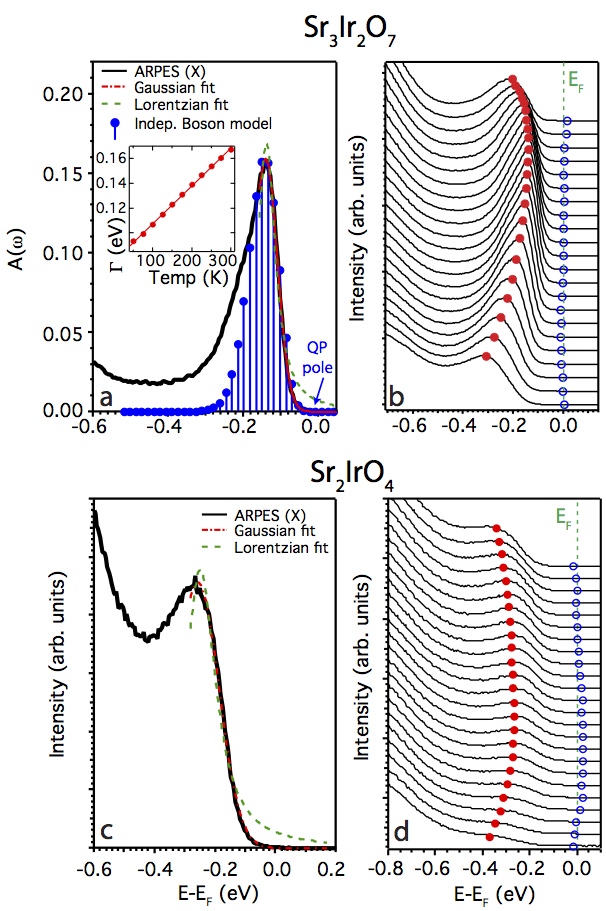}
\caption{ \label{f:lineshape} (a) EDC at the $X$-point (black), with the low-energy side of the peak fit by Gaussian (red dot-dashed) and Lorentzian (green dashed) lineshapes. An independent boson model calculation [$\omega_0=0.016$~eV, $g=6$, $T=50$~K, with the (zero-temperature) quasiparticle pole 30~meV below the chemical potential] is also shown, in excellent agreement with the measured ARPES data at low binding energies. The width of Gaussian fits to the low binding energy half of the peak in the spectral function of equivalent calculations as a function of temperature is shown inset. (b) EDCs close to the $X$-point. Red closed circles show the peak position of Gaussian fits, $E_B$. Blue open circles show $E_B+1.76\Gamma$, where $\Gamma$ is the width of the Gaussian fits. (c,d) Equivalent measurements (performed at 125~K to prevent charging) and fits for Sr$_2$IrO$_4$. In (d), blue open circles are shown at $E_B+1.5\Gamma$.}
\end{center}
\end{figure}

Instead, it appears to be driven by a substantial and continuous broadening of the spectral features with increasing temperature. Fig.~\ref{f:T-dep}(a) shows the temperature dependence of EDCs at the $X$-point. At temperatures as low as $\sim\!200$~K, the pronounced peak visible at lower temperatures has broadened significantly into a weak hump-like feature. Indeed, the linewidth of EDCs at both the $X$- (Fig.~\ref{f:T-dep}(b)) and $\Gamma$- (Fig.~\ref{f:T-dep}(c)) points both increase with temperature at a rate as high as 0.3~meV/K. This temperature evolution is almost identical for both warming and cooling of the same sample (Fig.~\ref{f:T-dep}(b,c)), ruling out sample ageing as the origin of the observed broadening. Rather, it reveals a significant temperature dependence of the many-body interactions, which is normally indicative of a strong coupling to the lattice. Furthermore, the spectral features we observe are distinctly non-quasiparticle like. Instead of a narrow Lorentzian peak at the top of the valence band~\cite{Baumberger:Phys.Rev.Lett.:96(2006)246402}, we consistently observe broad features that can only be satisfactorily fit using Gaussian lineshapes (Fig.~\ref{f:lineshape}(a)). 

Together, these observations are indicative of a polaronic ground state, driven by strong electron-boson coupling. In such a scenario, the quasiparticle residue becomes vanishingly small, and the photoemission measurements are dominated by incoherent excitations at higher binding energies that involve the simultaneous excitation of multiple bosons. Within a simple Franck-Condon broadening (FCB) picture, the measured spectral function is composed of distinct shake-off excitations separated by $\hbar\omega_0$, where $\omega_0$ is the bosonic mode frequency (Fig.~\ref{f:lineshape}(a)). In the solid state, these individual shake-off excitations are inherently broadened and so cannot be resolved. However, their intensity distribution, which reflects the overlap of the initial state with different excited final states, follows a Gaussian envelope, which is a hallmark of a polaronic system in photoemission measurements~\cite{Perfetti:Phys.Rev.Lett.:87(2001)216404,Shen:Phys.Rev.Lett.:93(2004)267002,Shen:Phys.Rev.B:75(2007)075115,Cardenas:Phys.Rev.Lett.:103(2009)046804,Tamai:Phys.Rev.B:74(2006)085407}. With increasing temperature, the linewidth of this Gaussian envelope broadens further, reflecting the presence of thermally-excited bosons~\cite{Mahan(2000)}. Our measured temperature-dependent linewidths are well described (Fig.~\ref{f:T-dep}(b,c)) by such a FCB model which was originally developed to describe photoemission spectra of the polaronic alkali-halides~\cite{Citrin:Phys.Rev.Lett.:33(1974)965}. From unconstrained fits of $\Gamma (T)$, we find a soft bosonic mode energy of $\omega_0\sim\!15$~meV and a strong coupling constant of $g\sim\!5-7$, which would place the system well within the strong coupling regime.

To further validate this picture, we employ these parameters for calculations within the independent boson approximation (IBA)~\cite{Mahan(2000)}. As shown in Fig.~\ref{f:lineshape}(a), the envelope of this model spectral function is in very good agreement with our measured photoemission spectra~\cite{Note2}. As for the experimental EDCs, the IBA calculations can only be adequately fit by Gaussian lineshapes. By explicitly including thermal population effects in our IBA calcualtions, we extracted the temperature dependence of such Gaussian envelopes, shown inset in Fig.~\ref{f:lineshape}(a), which increase almost linearly in width at a rate of 0.3~meV/K. This is in excellent agreement with our direct experimental measurements (Fig.~\ref{f:T-dep}(b,c)), and of the same order (although slightly smaller) than values observed in other polaronic systems~\cite{Citrin:Phys.Rev.Lett.:33(1974)965,Shen:Phys.Rev.B:75(2007)075115}. 

Moreover, our measured linewidths show a strikingly-similar momentum dependence to that of insulating cuprates~\cite{Shen:Phys.Rev.Lett.:93(2004)267002}. Away from the band top, the linewidth, $\Gamma$, increases directly proportional to the binding energy of the peak in the spectral function, $E_B$, such that $E_B(\mathbf{k})/\Gamma(\mathbf{k})=1.76\pm0.05$ (Fig.~\ref{f:lineshape}(b)). We find similar characteristics of the spectral features in Sr$_2$IrO$_4$, where we again observe broad Gaussian lineshapes (Fig.~\ref{f:lineshape}(c)) whose width is proportional to their binding energy (Fig.~\ref{f:lineshape}(d), $E_B(\mathbf{k})/\Gamma(\mathbf{k})=1.5\pm0.1$). Within the Franck-Condon picture, this implies that the quasiparticle poles have vanishing weight and almost no dispersion, which is consistent with observations in other polaronic systems~\cite{Shen:Phys.Rev.Lett.:93(2004)267002,Tamai:Phys.Rev.B:74(2006)085407}, and the findings of model calculations~\cite{Mishchenko:Phys.Rev.Lett.:93(2004)036402,Rosch:Phys.Rev.Lett.:95(2005)227002}. We also note that optical conductivity of Sr$_2$IrO$_4$ was found to show a pronounced dependence on temperature~\cite{Moon:Phys.Rev.B:80(2009)195110}, very similar to that of the insulating cuprate La$_2$CuO$_4$~\cite{Falck:Phys.Rev.Lett.:69(1992)1109}. Moreover, thermal excitation of polaronic carriers was recently proposed as the origin of the N{\'e}el transition~\cite{Kim:Phys.Rev.Lett.:109(2012)157402}. These observations all lend further support to our assignment of a polaronic ground state of insulating Ruddlesden-Popper iridates.

Our measured linewidths and binding energies of Sr$_2$IrO$_4$ are both larger than in the bilayer compound, reflecting a more robustly insulating ground state and stronger electron-boson coupling than in Sr$_3$Ir$_2$O$_7$. The microscopic origin of the bosonic mode in either compound, however, remains an open question. While it is tempting to ascribe it to a phonon due to the strong temperature dependence of the measured spectral features, the mode energies obtained from our fits are significantly lower than the dominant phonon branches~\cite{Cetin:Phys.Rev.B:85(2012)195148}. 
The strong spin-orbit coupling of these compounds is further expected to effectively couple lattice and spin excitations. Irrespective of this, our observations indicate that the quasiparticle poles are located much closer to the chemical potential than would be assumed from conventional interpretations of photoemission spectra within a weakly-interacting band picture. Using a simple calculation based on the Franck-Condon scheme proposed here and assuming particle-hole symmetric spectra, we can additionally reproduce the $\alpha$ peaks measured in optical conductivity~\cite{Kim:Phys.Rev.Lett.:101(2008)076402,Moon:Phys.Rev.Lett.:101(2008)226402} of both Sr$_3$Ir$_2$O$_7$ and Sr$_2$IrO$_4$. This further supports our model, and confirms that the spectroscopic band gap is much larger than the true quasiparticle gap, indicating that magnetic and/or lattice fluctuations play an essential role in stabilizing the insulating nature of layered $5d$ iridates.

\

This was was supported by the ERC, the UK EPSRC, the AFOSR (FA9550-12-1-0335) and the Grant-in-Aid for Scientific Research (S) (Grant No. 24224010). The research was carried out in part at the Stanford Synchrotron Radiation Lightsource, a Directorate of SLAC National Accelerator Laboratory and an Office of Science User Facility operated for the U.S. Department of Energy Office of Science by Stanford University.

\bibliographystyle{apsrev}

\end{document}